\documentclass{IEEEtran}
\usepackage{cite}
\usepackage{hyperref}
\usepackage{amsmath,amssymb,amsfonts}
\usepackage{algorithmic}
\usepackage{graphicx}
\usepackage{textcomp}
\usepackage{siunitx}
\def\BibTeX{{\rm B\kern-.05em{\sc i\kern-.025em b}\kern-.08em
    T\kern-.1667em\lower.7ex\hbox{E}\kern-.125emX}}
    \usepackage{color}

    \newcommand{\jj}{\mathrm{j}}
\newcommand{\E}{\mathrm{e}}

\newcommand{\comment}[1]{}

\newcommand{\uz}{\mathbf{\hat{z}}}

\newcommand\abs[1]{\left|#1\right|}

\begin{document}
\bstctlcite{IEEEexample:BSTcontrol}
\title{Symmetry and Finite-size Effects in Quasi-optical Extraordinarily THz Transmitting Arrays of Tilted Slots}
\author{Miguel Camacho, \IEEEmembership{Member, IEEE}, Ajla Nekovic, Suzanna Freer, Pavel Penchev, Rafael R. Boix, \IEEEmembership{Member, IEEE}, Stefan Dimov and Miguel Navarro-C\'ia, \IEEEmembership{Senior Member, IEEE}
\thanks{The work of M. Camacho was supported by the Engineering and Physical Sciences Research Council (EPSRC) of the United Kingdom, via the EPSRC Centre for Doctoral Training in Metamaterials [Grant No. EP/L015331/1]. The work of S. Freer was supported by the EPSRC [Studentship No. 2137478]. The work of R.R. Boix was supported by the Ministerio de Ciencia, Innovaci\'on y Universidades [Grant TEC2017-84724-P]. The work of M. Navarro-C\'ia was supported by the EPSRC [Grant No. EP/S018395/1], the Royal Society [Grant No. RSG/R1/180040], and the University of Birmingham [Birmingham Fellowship]. (\textit{Corresponding author: Miguel Navarro-C\'ia})}
\thanks{M. Camacho is with the Department of Electrical and Systems Engineering, University of Pennsylvania, Philadelphia, PA 19104-6390, USA (e-mail: mcamagu@seas.upenn.edu). }
\thanks{Ajla Nekovic is with the Faculty of Electrical Engineering, University of Sarajevo, Sarajevo 71000, Bosnia and Herzegovina}
\thanks{R. R. Boix  is with Microwaves Group, Department of Electronics and Electromagnetism, College of Physics, University of Seville, Avda. Reina Mercedes s/n, 41012, Seville, Spain  (e-mail: boix@us.es)}
\thanks{S. Freer and M. Navarro-C\'ia are with the School of Physics and Astronomy, University of Birmingham, Birmingham B15 2TT, United Kingdom (e-mail: M.Navarro-Cia@bham.ac.uk)}
\thanks{P. Penchev and S. Dimov are with the Department of Mechanical Engineering, School of Engineering, University of Birmingham, Birmingham, B15 2TT, UK}}

\maketitle

\begin{abstract}
Extraordinarily transmitting arrays are promising candidates for quasi-optical (QO) components due to their high frequency selectivity and beam scanning capabilities owing to the leaky-wave mechanism involved. We show here how by breaking certain unit cell and lattice symmetries, one can achieve a rich family of transmission resonances associated with the leaky-wave dispersion along the surface of the array. By combining two dimensional and one dimensional periodic Method of Moments (MoM) calculations with QO Terahertz (THz) time-domain measurements, we provide physical insight, numerical and experimental demonstration of the different mechanisms involved in the resonances associated with the extraordinary transmission peaks and how these evolve with the number of slots. Thanks to the THz instrument used, we are also able to explore the time-dependent emission of the different frequency components involved.  
\end{abstract}


\section{Introduction}
\label{sec:introduction}
\IEEEPARstart{I}{n} the early 1990s, the extraordinary optical tranmission (EOT) phenomenon through subwavelength apertures was discovered \cite{Betzig1986,Betzig1988,Ebbesen1998} opening the door to new and exciting physics. The transmission peak frequency and magnitude were initially thought to invalidate Bethe's prediction for subwavelength apertures \cite{Bethe44}, although this comparison lacked a very meaningful feature of the experimental sample: the periodicity. This characteristic held the key for the explanation of the, at that point, unexplained mystery.

The relation between the high transmissivity and the periodicity became apparent when the transmission spectrum was mapped for different planes of incidence \cite{Ghaemi98}, finding that these peaks correspond to the excitation of optical-frequency surface waves known as surface plasmons \cite{Ritchie1957, Barnes2003}. These surface waves, usually tightly bound to the surface of a plasmonic material, can become leaky when supported by periodic structures, following the same principle used nowadays by periodically-modulated leaky wave antennas \cite{Minatti2018} and that can be traced back to \cite{Oliner1959}. This intuitive explanation was later confirmed by means of plane wave expansion \cite{Martin-Moreno2001}, showing that these surface waves are capable of providing large transmission enhancements even for screens with non-negligible thicknesses thanks to the evanescent coupling between surface plasmons supported by opposite faces.

\begin{figure}[!t]
\centerline{\includegraphics[width=\columnwidth]{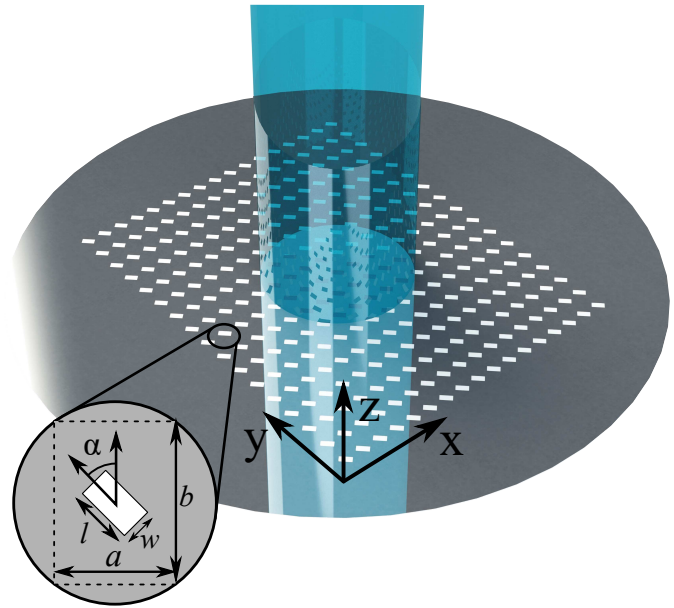}}
\caption{Schematic diagram of the freestanding subwavelength tilted slot $N_x$ by $N_y$ array along with the collimated incident Gaussian beam. Lattice periods \textit{a} and \textit{b}, and slot dimensions $l\times w$ rotated by an angle $\alpha$.}
\label{fig1}
\end{figure}

Experiments similar to those published in \cite{Ebbesen1998} were conducted at microwave regime, in the absence of plasmonic materials, shockingly finding similar results \cite{Miyamuru04,Cao2004,Beruete05,Medina2009}. Also, EOT was reported at such regimes for s-polarized illumination that cannot excite surface plasmons even if plasmonic materials were used \cite{Kuznetsov2009a,Beruete2011b}. Without plasmons, the explanation of microwave extraordinary transmission was challenging. However, not one but two complementary explanations were provided. The first one followed the same logic as that one provided at optical frequencies: although metals do not support plasmons at microwave frequencies, when properly geometrically engineered, they can support analogous surface wave modes \cite{Pendry2004,Hibbins2004}, which produce extraordinary transmission effects analogous to those of surface plasmons \cite{Lomakin2006}. It is worth noting that although these surface waves received a large attention subsequently to their role in EOT, they were known by the engineering community since the mid twentieth century \cite{Cutler,Walter1970}. An alternative explanation, based on impedance matching formulation well known for transmission lines was provided over a decade ago \cite{Medina2008, Beruete2011a}.

The development of extraordinary transmission was subsequent to similar periodic structures, widely-used by the microwave community, known as Frequency Selective Surfaces (FSS) \cite{Munk2000, Mittra88}. However, the latter make use of the \emph{natural} resonance of the unit element in each until cell, defined this as the one dictated by their electrical size, with little or no effects arising from their arrangement \cite{Rodriguez-Ulibarri2017}. This main difference is the root of interesting phenomena, as imposing or breaking certain symmetries has drastic effects on the existence of EOT peaks associated with the periodicities of the system. For instance, when the unit cell presents a symmetry plane aligned with the periodicity (electric wall, for instance), this inhibits the EOT peak associated with such lattice vector \cite{Camacho2017e}. As it was shown there, this means that when finite arrays are considered, one can ignore the finite size effects along one of the periodicities if the unit cell presents symmetries that inhibit that resonance, as the surface wave along the short direction will not couple to the radiation and therefore cannot sense the truncations of the system \cite{Camacho2019b, Camacho2019}.

EOT arrays, with their narrow linewidth and strong suppression out of band, have been shown to be excellent candidates for quasi-optical (QO) components such as filters and wave plates \cite{Kuznetsov2009a,Beruete2007b,Torres2014}. However, the potential added by the exploitation of the (lack of) symmetries is still to be explored. In an effort to bridge this gap, in this paper we study the extraordinary transmission phenomenon through arrays of non-symmetric slot arrays. This lack of symmetry is two-fold: we make use of rectangular arrays (with periodicities given by $a$ and $b$, with $a$ different from $b$), whose unit cell contains slots that present a rotation with respect to the lattice vectors and also insert a slot-shaped aperture which is not colinear with any of the lattice vectors. In the first part of the paper we show how these two conditions are required for us to explore the whole set of resonances associated with the two different periodicities, and in the second part we experimentally demonstrate the presence of these resonances, and show how finite effects also play an important role when realistic collimated illuminations are used. Finally we explore both the time-response of the system through spectrography and illumination effects such as polarization and angle of incidence dependencies.



\section{Theory} \label{sectheory}

Let us consider the array shown in Fig.\ref{fig1}, which contains an array of slots of dimensions $l\times w$, rotated with respect to the two periodicities, with values $a$ and $b$. These slots are cut into an infinite perfectly-conducting zero-thickness screen, which is a valid approximation for terahertz (THz) frequencies, such as those studied here. This rectangular geometry for the apertures is very convenient for the semi-analytical solution of the scattering problem in terms of analytical basis functions, yet the phenomena presented in the following can be found for any other geometry that lacks mirror symmetry. Let us consider the case in which this array of slots is illuminated by a plane wave propagating along the $\uz$ direction, whose electric field is given by $\mathbf{E}_\text{i}$. The problem of the scattering by the array can be studied by means of an integral equation for the tangential electric field in the slots, 
$\mathbf{E}_t^{\mathrm{sc}}$, given by
\begin{figure*}
\centering
  \includegraphics[width=\textwidth]{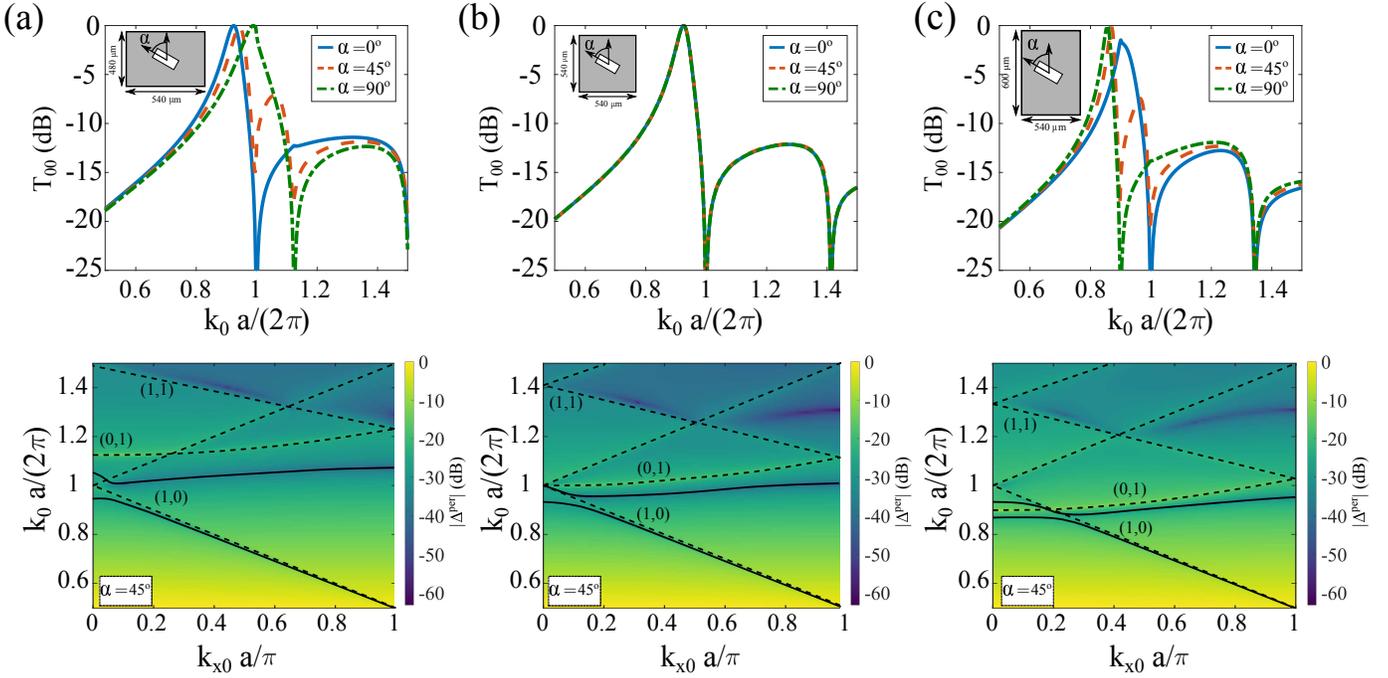}
  \caption{Top figures show the transmission spectra at normal incidence for the lowest Floquet mode for vertical, diagonal ($\alpha=45^\circ$) and horizontal slots, with the electric field polarization perpendicular to the long side of the slots, for (a) $a=\SI{540}{\micro\meter}$ and $b=\SI{480}{\micro\meter}$, (b) $a=b=\SI{540}{\micro\meter}$ and (c) $a=\SI{540}{\micro\meter}$ and $b=\SI{600}{\micro\meter}$. Bottom figures show the real part of the associated dispersion diagrams calculated for the diagonal slot ($\alpha=45^\circ$), over a color map representing a cut of the three dimensional complex space value of the MoM matrix determinant. In the dispersion diagrams, dashed lines represent the labelled lightline spatial harmonics. All the data in this figure was obtained for $l=\SI{220}{\micro\meter}$ and $w = \SI{175}{\micro\meter}$.}
 \label{fig2}
\end{figure*}
\begin{align}
\label{inteq1}
\nonumber
&\mathbf{J}^{\mathrm{as}}+\int_{-\infty}^{\infty}\int_{-\infty}^{\infty} \!
\overline{\mathbf{G}}_M(x-x',y-y') \\
&\cdot \mathbf{E}_t^{\mathrm{sc}} (x',y',z=0) \ dx' dy'=\mathbf{0}\qquad (x,y) \in \mathrm{slots}
\end{align}
\noindent where $\overline{\mathbf{G}}_M(x,y)$ is a dyadic Green's function, relating the in-plane components of the electric field on the surface of the slots and the surface current density on the metallic screen, and $\mathbf{J}^{\mathrm{as}}$ is the surface current density on the perfectly conducting plane in the absence of the slots (dictated by $\mathbf{E}_\text{i}$). It is well-known that for the case of a doubly-infinitely periodic array, the integral equation can be reduced to a single unit cell defining a periodic dyadic Green's function, $\overline{\mathbf{G}}_M^{\mathrm{per}}$ which groups the contributions from all the arrays through an infinite summation, in which the fields on the array are considered Floquet-periodic as imposed by the impinging wave \cite{Camacho2016}. The periodic expression of \eqref{inteq1} is then given by
\begin{align}
\label{inteq2}
\nonumber
&\mathbf{J}^{\mathrm{as}}+\int_{0}^{a}\int_{0}^{b}
\!\overline{\mathbf{G}}_M^{\mathrm{per}}(x-x',y-y') \\
&\cdot \mathbf{E}_t^{\mathrm{sc}} (x',y',z=0) \ dx' dy'=\mathbf{0} \qquad (x,y) \in \delta_{00}
\end{align}
\noindent where the integral is limited to the single unit cell $\delta_{00}$ by defining a two-dimensionally periodic Green's function given by
\begin{equation}
\label{perGr}
\overline{\mathbf{G}}_M^{\mathrm{per}}(x,y)=\sum_{m,n=-\infty}^{+\infty}
\overline{\mathbf{G}}_M(x-ma,y-nb)\ \E^{\jj( k_{x0}ma + k_{y0}nb)}
\end{equation}
where $k_{x0}$ and $k_{y0}$ are the phasing wavevectors defined as in \cite{Camacho2017e}. 

Both equations \eqref{inteq1} and \eqref{inteq2} can be solved by means of the Method of Moments (MoM) \cite{Harrington93}, which was used in \cite{Camacho2019b} to tackle the problem of periodic non-rotated slots, and therefore will be just summarized here. In this method, the unknown tangential electric field $\mathbf{E}_t^{\mathrm{sc}}$ is expanded as a linear combination of proposed basis functions. When this linear expansion is introduced into either \eqref{inteq1} or \eqref{inteq2}, one can build a system of linear equations for the weights of those basis functions, that can be readily solved by numerical means. The key of this method is the use of basis functions that can render the features of the fields in the slot aperture, reproducing the singular behaviour of the fields near the edges for both polarization. Previous studies show that only a very reduced set of them is needed (5 at most) when these are chosen to be Chebyshev polynomials weighted by the singular square-root singular behaviour of the field for each polarization \cite{Camacho2017b,Lerer93,Camacho2019a}. 

Once the system of equations has been inverted, the total electric field on the surface of the array can be directly calculated together with its Fourier transform, which can be used to calculate the far-field and the scattering parameters \cite{Camacho2016}.

As discussed in the introduction, the presence of EOT phenomenon is linked to the existence of surface waves supported by the array. Mathematically this link is clear, as the existence of surface waves can be studied by looking for the zeroes of the determinant of the matrix of the system of linear equations resulting from the application of the MoM, $\abs{\Delta^\text{per}}$, which is a function of the geometrical parameters, the frequency \emph{and} the phasing imposed between adjacent unit cells. This means that when the excitation is close in frequency, phasing and polarization to that of the surface wave, resonant behaviour is found. When the position of these zeroes is tracked for a range of frequencies and phasings, one can build the dispersion diagram of the surface modes \cite{Camacho2017a}.

Using the analysis techniques and the concepts presented here, we have studied the transmission through a two dimensional periodic array of slots for three different periodicity configurations, i.e. $a>b$, $a=b$ and $a<b$, as shown in Fig. \ref{fig2}(a-c), which contain both the transmission spectra at normal incidence (top figures) and the dispersion diagram of the leaky surface waves supported by the array along the $x$ direction for the case of 45$^\circ$ slot rotation. Fig. \ref{fig2} has been obtained for a doubly infinite periodic array of slots (i.e., through the solution of Eqn. \eqref{inteq2} with a periodic dyadic Green's function).

In Fig. \ref{fig2}(b), the well-studied EOT through a square slot array is presented for benchmarking purposes, for which we present both transmission spectra for three different rotation angles at normal incidence and the dispersion diagram for the 45$^\circ$ slot rotation. Given the equality between the lattice vectors associated with both periodicities, their EOT resonances appear at the same frequency, and the system has a single EOT peak as the screen zero-thickness inhibit the so-called odd mode that appears very close to the frequency of the first Wood-Rayleigh's anomaly \cite{Medina2008}. In terms of the surface wave dispersion shown in the bottom figure, one can see only one mode crosses the broadside direction.

When the periodicity along the $y$ direction is smaller than that along the $x$, the degeneracy of the EOT peak is broken, and we would expect to find two peaks in the transmission spectrum in Fig. \ref{fig2}(a). However, when the slot is aligned with either of the periodicities and the impinging electric field is perpendicular to that, one see that the peak associated with the lattice vector parallel to the slot disappears. In contrast, both EOT peaks are present for any intermediate value. The fact that the existence of one of the periodicities can be completely hidden through the use of symmetries is remarkable, and its explanation can be found both in the mathematics and in the physics of the problem. Mathematically, the EOT is linked with the pole-type divergence of the Green's function at the onset of a diffraction lobe. However, the symmetry of the problem can lead to a pole-zero cancellation when the scattered electric field points perpendicularly to the in-plane wavevector of the diffraction mode that is generating such pole divergence, via a zero in the diagonal term of the dyadic Green's function \cite{Camacho2017e}. From the point of view of the surface wave modes responsible for the EOT phenomenon, this absence of EOT peaks under certain illuminations can be explained in terms of polarization mismatch between the impinging radiation and the modes supported by the array. Note that freestanding connected conducting structures, such as arrays of apertures, support transverse magnetic (TM) modes, while disconnected conducting surfaces such as arrays of patches, support transverse electric (TE) modes. Therefore, if one wants to excite the surface mode propagating along the $x$ direction, one would need to be able to excite a magnetic component along the $y$ direction, and \textit{vice versa}. In the presence of symmetry (i.e., slot parallel to one of the axes), the impinging polarization across or along the slot axis will be preserved, and only one of the two modes will be excited. Therefore, the EOT peak associated with the not-excited mode will be suppressed. However, if one considers a slot that is not aligned with the axes, the slot field distribution will couple to both surface waves, therefore exciting both EOT peaks.

In Fig. \ref{fig2}(c) we present the case in which the EOT associated with the periodicity along $y$ appears at a frequency lower than that associated with the periodicity along $x$. As seen earlier for Fig. \ref{fig2}(a), when the slot is aligned with either of the axes, the periodicity along the direction parallel to the long sides of the slot will not be present in the transmission spectrum, although its presence can be noticed due to the loss in transmission of the zero-th order Floquet (see the solid blue line of the top Fig. \ref{fig2}(c)), due to the interesting fact that although the EOT peak does not appear, its associated Floquet mode can couple to the energy of the scattered fields once it is above its cut-off frequency, while avoiding the singularity of its onset. In terms of the dispersion diagram, in the bottom Fig. \ref{fig2}(c), one can see that the two modes associated with the surface waves propagating along $x$ and $y$ are pushed together by their corresponding lightlines, giving place to a mode coupling through a Morse critical point \cite{Yakovlev1997}.

The fact that the position of the two resonant peaks can be dictated by the periodicities chosen along the $x$ and $y$ direction allows for a large freedom in the design of quasi-optical filters. The electrical size of the aperture can be then used to engineer the width of the resonances, as done in standard single-resonance EOT filters \cite{Camacho2016}. In the particular case of tilted slots, the angle $\alpha$ can be used for the design of the transmission levels of the two peaks, as it tunes the coupling between the scattered field and the two surface wave families, each associated to one periodicity.

As one can see, the EOT through tilted slot arrays opens a wide variety of situations through the coupling of the impinging radiation into a wider variety of Floquet modes, and therefore of EOT resonances. As it was shown in \cite{Camacho2019c}, for the cases of highly symmetrically positioned slots, the finite size along the non-excited EOT resonance can be disregarded, however, this is not the case when one considers tilted slots. In the following let us focus on the truncation effects that have to be considered for any realistic implementation of quasi-optical filters based on this multi-EOT phenomenon.

\section{Measurements and Discussion} \label{secexperiment}
In order to verify the theoretical results and physical insight provided in the previous section, we performed a series of experiments at THz frequencies using QO experimentation techniques as discussed in the Appendix. The samples under test were laser micromachined (further details about the fabrication can also be found in the Appendix) subwavelength hole arrays on 10$\mu$m-thick aluminium foils. The geometrical dimensions of the fabricated samples were inspected with a microscope; the in-plane lattice periods were $\mathrm{\textit{a} = 540 \pm \SI{5}{\micro\meter} }$ and $\mathrm{\textit{b} = 600 \pm \SI{5}{\micro\meter}}$, and the slot dimensions were $l = 226  \pm \SI{5}{\micro\meter}$, $w = 176 \pm \SI{5}{\micro\meter}$.

\begin{figure}[!t]
\centerline{\includegraphics[width=0.9\columnwidth]{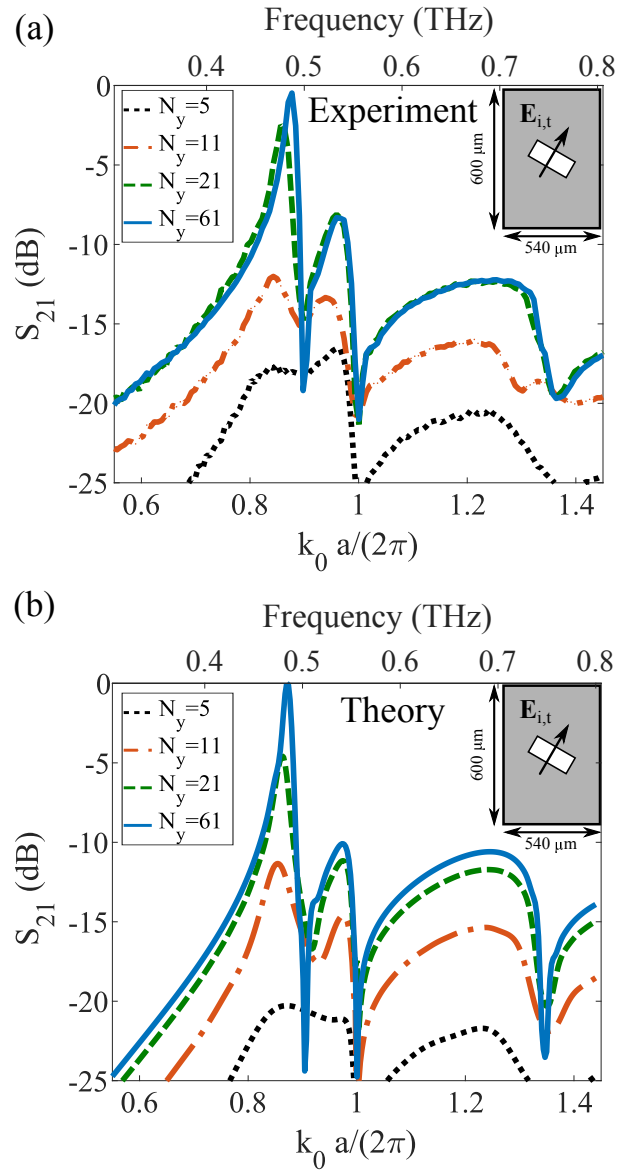}}
\caption{On-axis transmission through freestanding arrays of diagonal ($\alpha=45^\circ$) slots and varying number of rows $N_y$ with lattice periods $\mathrm{\textit{a} \approx \SI{540}{\micro\meter}}$ and $\mathrm{\textit{b} \approx \SI{600}{\micro\meter}}$, and slot dimensions $l \approx \SI{220}{\micro\meter}$ and $w \approx \SI{175}{\micro\meter}$: measurements (a) and MoM results (b).}
\label{fig3}
\end{figure}

\begin{figure*}[htpb]
\centering
  \includegraphics[width=\textwidth]{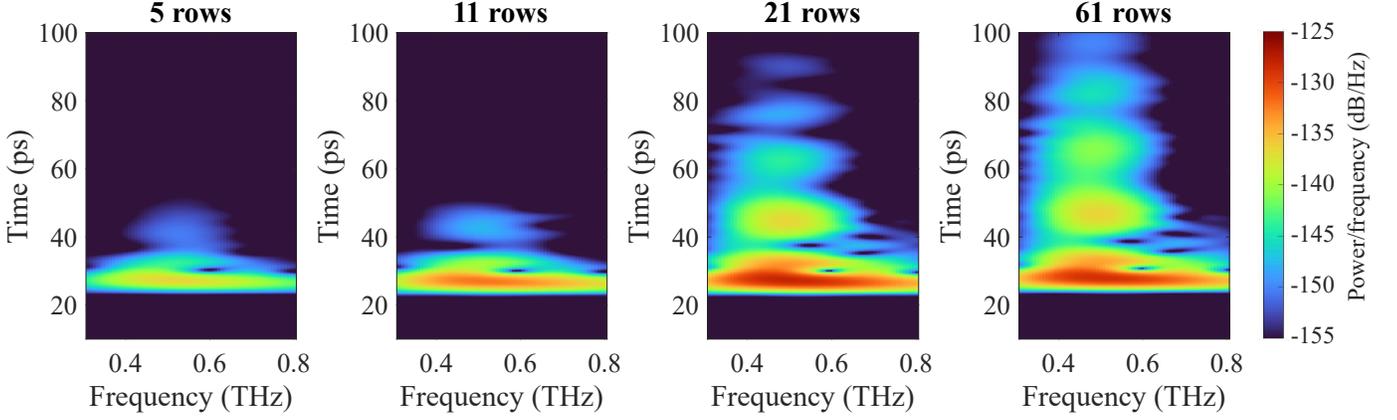}
  \caption{From left to right, spectrogram (i.e. time-frequency-maps) of the detected waveform (co-polar measurement) for samples with $N_y$ = 5, 11, 21 and 61 when the incident electric field is polarized perpendicularly to the long side of the slots.}
  \label{fig4}
\end{figure*}

The transmission spectrum at normal incidence with the electric field polarization perpendicular to the long side of the slots, $l$, increases with the number of rows $N_y$ (Fig. \ref{fig3}) for diagonal slots. Notice that $N_x \geqslant 65$ across the whole wafer in all fabricated samples, and thus, the truncation along $x$ direction was assumed to have no impact in the electromagnetic response. The MoM approach described in [35] can be then used for the very efficient analysis of the structures involved in the experiments by reducing the problem to a single strip of slots which is repeated periodically along a single direction. Under this approach, integral equation in \eqref{inteq1} becomes
\begin{align}
\nonumber
&\mathbf{J}^{\mathrm{as}}(x,y)+\sum_{j=0}^{N_y-1} \int\!\!\!\int_{\eta_{0j}} \!
\overline{\mathbf{G}}_{\mathrm{1D}}^{\mathrm{per}}(x-x',y-y') \\
\label{inteq1d}
&\cdot \mathbf{E}_t^\text{sc} (x',y',z=0) \ dx' dy'=\mathbf{0} \quad (x,y) \in \eta_{0j}\\
\nonumber
& \qquad\qquad\qquad (j=0,\dots,N_y-1),
\end{align}
where $\eta_{0j}$ represents the surface of the j-th slot within the unit strip and $\overline{\mathbf{G}}_{\mathrm{1D}}^{\mathrm{per}}(x,y)$ is the 1-D periodic dyadic Green's function given in terms of the free space Green's function $\overline{\mathbf{G}}_M(x,y)$
\begin{equation}
\label{eqG1D}
\overline{\mathbf{G}}_{\mathrm{1D}}^{\mathrm{per}}(x,y)=\sum_{i=-\infty}^{+\infty}
\overline{\mathbf{G}}_M(x-ia,y)\E^{\jj ik_{x0}a}.
\end{equation}

To introduce the illumination effects, a 1-D Gaussian beam profile is used as the impinging wave for the calculation of $\mathbf{J}^{\mathrm{as}}(x,y)$, as in \cite{Camacho2019b}. In the experiment, almost total transmission is obtained when $N_y = 61$, whereas the MoM predicts total transmission. The slight quantitative disagreement between the results for this $N_y$ is arguably, to a large extent, due to the absence of loss (i.e. ohmic and scattering losses due to surface imperfections) in the MoM, and, to a lower extent, due to the two-dimensional MoM calculation in which the incident Gaussian beam has a beam-waist of 4 mm along $y$ and infinite along $x$. For such array with $N_y = 61$, both the direct transmission and the contribution of the surface mode propagating along the $y$ direction are saturated. Indeed, saturation of the direct transmission actually happens already for $N_y = 21$ given the beam-waist of the Gaussian beam.

The existence of two transmission channels involved in the EOT (i.e. direct transmission and leaky-wave-mediated transmission) becomes evident in the time-frequency analysis of the scattering phenomenon, which allows for the time-dependent analysis of the radiated waves \cite{Freer2019}. To this end, Fig. \ref{fig4} shows the spectrograms for the different truncated arrays of diagonal slots with varying number of slots along the $y$ direction, $N_y$. Regardless of $N_y$, we find that a significant amount of the energy arriving at the detector does it around 28 ps across the whole spectral window. This is associated with the contribution of the direct transmission, as the whole-array standing wave has not had enough time to form \cite{Camacho2016}. Meanwhile, there is a prolonged presence of energy arriving beyond 28 ps only at the frequency of the EOT peak. The larger $N_y$ is, the longer this ringing of energy lasts. This is due to the fact that more elements are involved in the resonance, increasing its quality factor (and therefore the frequency selectivity as shown in Fig.~\ref{fig3}). This resonance is formed as the superposition of counter-propagating leaky waves, which are excited at a precise frequency, the EOT frequency, dictated by the leaky-wave complex dispersion relation. Such a slow build-up of the resonance is possible due fact that the crossing of the leaky-wave dispersion with the broadside direction occurs at a zero slope (and therefore zero group velocity). As it was shown in \cite{Navarro-Cia2018a,Camacho2019b}, when Gaussian beams are considered, the transmission peak is slightly red-shifted, meaning that the resonance is mediated by slowly-propagating waves, but not completely static ones, contrarily to that found when using plane waves in infinite periodic arrays.

\begin{figure}[!t]
\centerline{\includegraphics[width=0.9\columnwidth]{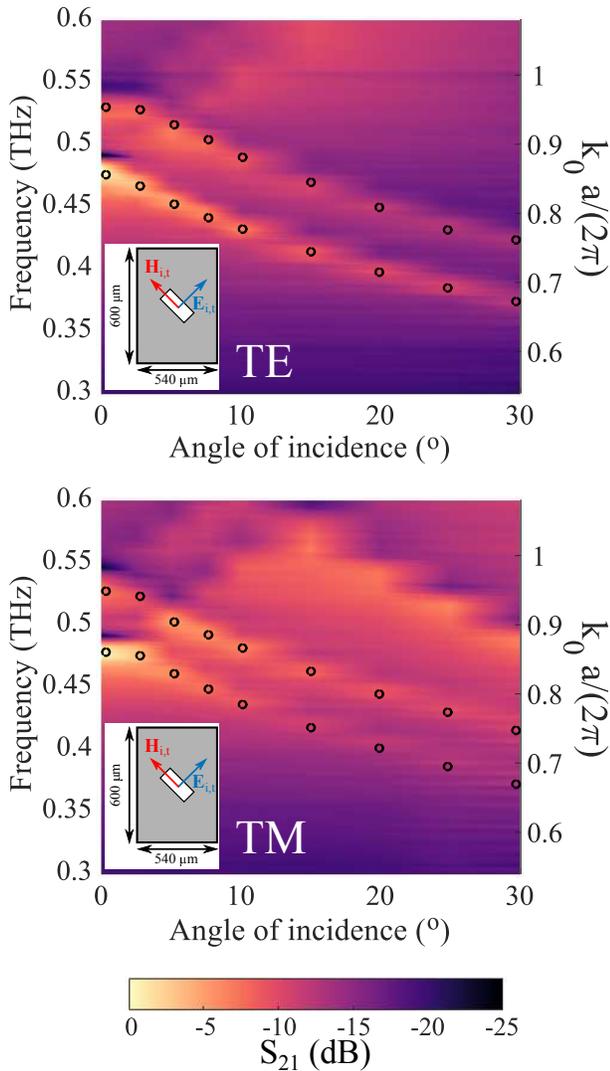}}
\caption{Transmission spectrum as a function of the angle of incidence for TE (rotation of the sample along the \textit{E}-axis) and TM polarization (rotation of the sample along the \textit{H}-axis), with black circles marking the largest transmission for each angle of incidence. The electric field is perpendicular to the long side of the diagonal ($\alpha=45^\circ$) slots.}
\label{fig5}
\end{figure}

We have also studied the complete angular dependence of the system by varying both angle of incidence and polarization, as these allow to control the matching of the impinging wave to the two leaky waves responsible for the two different EOT resonances of the tilted slot array.

Fig.~\ref{fig5} shows the angle of incidence dependence of the measured transmission spectrum under both TE and TM wave illumination, whose planes of incidence are defined by the $z$ direction and the vectors shown in the insets of the figure. In contrast to what we find when exploring the dispersion along the $x$ axis in Fig.~\ref{fig2}, at both planes of incidence associated with TE and TM incidence, the two lightlines associated with the (1,0) and (0,1) harmonics show very similar dispersion relation, enforcing a similar angle-dependence on the EOT peaks. Additionally, one can see the excitation of higher order spatial harmonics, such as the (-1,0) and (0,-1), whose associated resonant frequencies increase with the angle of incidence. Although for the sake of brevity we only show here the position of the peaks, we have found that the relative transmissivity of the two EOT peaks can be controlled by choosing the wave to be either TE or TM, therefore promoting the excitation of one of the two leaky waves while suppressing the other. In particular, by using TE oblique incidence the lower frequency resonance is favored, while the TM oblique incidence enhances the higher EOT resonance. This effect can be explained by the increasing wavevector matching between the impinging and the leaky wave, which can ultimately lead to huge resonant amplitudes as theoretically shown for bound surface waves supported by semi-infinite arrays \cite{Camacho2019}.

\begin{figure}[!t]
\centerline{\includegraphics[width=0.9\columnwidth]{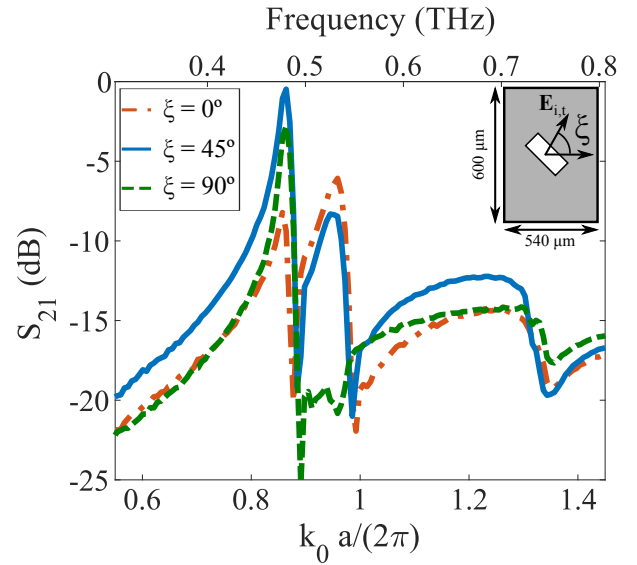}}
\caption{Measured on-axis transmission through  freestanding  arrays of diagonal slots for different rotation of the $N_y$ = 61 sample along the $z$-axis as indicated by the inset in the figure.}
\label{fig6}
\end{figure}

We have found experimentally, however, that the excitation of the two leaky waves can also be controlled by playing with the field configuration of the scattered fields on the slot. In Fig.~\ref{fig6}, we show the normal incidence transmission coefficient for varying angles of electric field polarization. We find that the higher frequency EOT, associated with the periodicity along the $x$ direction can be almost completely suppressed when the impinging excitation is polarized along the $y$ direction. This is due to the modification of the field distribution, as in contrast with all the cases studied before, now there is a non-negligible electric field component excited along the long side of the slot, which allow for the coupling with a larger number of modes.


\section{Conclusion} \label{seconclusion}
In this paper we have provided an extensive study of the EOT phenomenon through periodic and finite arrays of tilted slots, which allow for a very distinguishable simultaneous coupling to two different EOT resonances, in contrast to the highly symmetric arrays traditionally studied in the literature. We have shown that the two resonances can be linked to the existence of two different leaky waves, propagating along perpendicularly to each other, as demonstrated by the calculation of their dispersion diagrams and also by showing their matching (or absence of matching) to the scattered fields under certain symmetries. By performing experiments at terahertz frequencies we have both demonstrated the validity of our theory, but have also shown the role that finite effects play in the quality factor of the theoretically highly selective EOT resonance. Thanks to the time-domain spectroscopy, we are able to show the consequences that the dispersion characteristics of the leaky-waves involved have in the time response associated with each frequency component of the scattered fields, namely its slow resonance build-up. By exploring the angular dependence under two differently polarized illuminations, we have shown the off-axis dispersion of the leaky waves which is dominated by the linear lightline harmonic dispersion and also the ability to suppress or enhance the EOT resonances via off-axial excitation of the slots. The insight provided here sets the design guidelines for advanced quasi-optical components based on extraordinary transmission phenomenon.

\section*{Appendix}
\textit{Fabrication}. 
The laser fabrication trials were performed on a state-of-art laser micromachining system that integrates a femtosecond (fs) laser source (Amplitude Systems Satsuma model) with a maximum average power of 5W, central emission wavelength of 1030 nm, pulse duration of 310 fs and maximum pulse repetition rate of 500 KHz. A stack of five mechanical axes (three linear and two rotary stages) were used to move the samples with very high accuracy and precision, i.e. the positioning resolutions of the linear and rotary stages as stated by the manufacturer were 0.25 $\mu$m and 45 $\mu$rad, respectively \cite{Worldwide2010}. The laser system was also equipped with a high dynamics 3D optical scan head for structuring with scanning speeds up to 2 m/s and thus to obtain relatively high material ablation rates. A 100 mm focal length telecentric lens is mounted at the exit of the beam path to achieve a focused beam spot diameter of 30 $\mu$m. The achievable processing accuracy and repeatability with the 3D scan head is better than $\pm10 \mu m$ for all scanning speeds used in the experiments, while its processing volume is 30 mm (X) $\times$ 30 mm (Y) $\times$ 10 mm (Z) \cite{Penchev2017}. The specific laser parameters used to produce the filters are detailed in Table 1 below.

\begin{table}[]
\centering
\caption{Laser parameters for the fabrication of the finite size tilted slot arrays.}
\begin{tabular}{|l|l|l|}
\hline
\textbf{Laser Parameter} & \textbf{Unit} & \textbf{Value} \\ \hline
Power                    & W              & 3.9            \\ \hline
Frequency                & kHz            & 500            \\ \hline
Pulse energy             & $\mu$J             & 7.8            \\ \hline
Scanning speed           & m/s            & 2              \\ \hline
Pulse duration           & fs             & 310            \\ \hline
Beam spot diameter       & $\mu$m             & 30             \\ \hline
Laser beam polarization  & -              & Circular       \\ \hline
Hatch style              & -              & Random         \\ \hline
Hatch pitch              & $\mu$m             & 4              \\ \hline
Ablation rate per layer  & $\mu$m             & 1              \\ \hline
\end{tabular}
\end{table}

\textit{Measurements}. The samples were characterized with the all fibre-coupled THz time-domain spectrometer TERA K15 from Menlo Systems in a QO configuration without purging. A couple of TPX planoconvex lenses (effective focal length $\approx$54 mm) were used to work with a collimated beam whose frequency-dependent beam-waist at the sample position was estimated to be 4.1 mm and 4.0 mm along the \textit{E}- and \textit{H}-plane at 0.5 THz, respectively. For the angle-resolved measurements of Fig.~\ref{fig5} (with an angular step of 2.5$^\circ$ until 10$^\circ$ and 5$^\circ$ until 30$^\circ$) and Fig.~\ref{fig6}, the $N_y$ = 61 sample was installed on a manual rotation stage and on a high-precision rotation mount. The lock-in constant was set to 300 ms for the angle-resolved measurements of Fig.~\ref{fig5} and 100 ms for the rest, whereas the total temporal length of the recorded waveforms was at least 260 ps to have a spectral resolution of 3.8 GHz in the worst case. Calibration was done by comparing the measurements with the line-of-sight configuration (i.e. emitter and detector on-axis) without the sample on the sample holder. To generate the spectrograms (i.e. short-time Fourier transform) in Fig. \ref{fig4}, each waveform is divided into segments of length 8.9 ps and an overlap of 98\%, which are windowed with a Hann window.

\section*{Acknowledgements}
The authors would like to thank Dr. J. Churm (University of Birmingham) for measuring the unit cell dimensions of the fabricated truncated arrays.

\bibliography{library}

\begin{thebibliography}{10}
\providecommand{\url}[1]{#1}
\csname url@samestyle\endcsname
\providecommand{\newblock}{\relax}
\providecommand{\bibinfo}[2]{#2}
\providecommand{\BIBentrySTDinterwordspacing}{\spaceskip=0pt\relax}
\providecommand{\BIBentryALTinterwordstretchfactor}{4}
\providecommand{\BIBentryALTinterwordspacing}{\spaceskip=\fontdimen2\font plus
\BIBentryALTinterwordstretchfactor\fontdimen3\font minus
  \fontdimen4\font\relax}
\providecommand{\BIBforeignlanguage}[2]{{%
\expandafter\ifx\csname l@#1\endcsname\relax
\typeout{** WARNING: IEEEtran.bst: No hyphenation pattern has been}%
\typeout{** loaded for the language `#1'. Using the pattern for}%
\typeout{** the default language instead.}%
\else
\language=\csname l@#1\endcsname
\fi
#2}}
\providecommand{\BIBdecl}{\relax}
\BIBdecl

\bibitem{Betzig1986}
E.~Betzig, A.~Lewis, A.~Harootunian, M.~Isaacson, and E.~Kratschmer, ``{Near
  Field Scanning Optical Microscopy (NSOM): Development and Biophysical
  Applications},'' \emph{Biophys. J.}, vol.~49, no.~1, pp. 269--279, 1986.

\bibitem{Betzig1988}
R.~E. Betzig, ``{Nondestructive Optical Imaging of Surfaces with 500 Angstrom
  Resolution.}'' Ph.D. dissertation, Cornell University, 1988.

\bibitem{Ebbesen1998}
T.~W. Ebbesen, H.~J. Lezec, H.~F. Ghaemi, T.~Thio, and P.~A. Wolff,
  ``{Extraordinary optical transmission through sub-wavelength hole arrays},''
  \emph{Nature}, vol. 391, no. 6668, pp. 667--669, Feb. 1998.

\bibitem{Bethe44}
H.~A. Bethe, ``{Theory of diffraction by small holes},'' \emph{Phys. Rev.},
  vol.~66, no. 7-8, pp. 163--182, Oct. 1944.

\bibitem{Ghaemi98}
H.~F. Ghaemi, T.~Thio, D.~E. Grupp, T.~W. Ebbesen, and H.~J. Lezec, ``{Surface
  plasmons enhance optical transmission through subwavelength holes},''
  \emph{Phys. Rev. B}, vol.~58, no.~11, pp. 6779--6782, Sep. 1998.

\bibitem{Ritchie1957}
R.~H. Ritchie, ``{Plasma Losses by Fast Electrons in Thin Films},'' \emph{Phys.
  Rev.}, vol. 106, no.~5, pp. 874--881, 1957.

\bibitem{Barnes2003}
W.~L. Barnes, A.~Dereux, and T.~W. Ebbesen, ``{Surface plasmon subwavelength
  optics},'' \emph{Nature}, vol. 424, no. 6950, pp. 824--830, 2003.

\bibitem{Minatti2018}
G.~Minatti, M.~Faenzi, M.~Mencagli, F.~Caminita, D.~Gonz, C.~D. Giovampaola,
  A.~Benini, E.~Martini, M.~Sabbadini, and S.~Maci, ``{Metasurface Antennas},''
  in \emph{Aperture Antennas Millim. Sub-millim. Wave Appl.}\hskip 1em plus
  0.5em minus 0.4em\relax Springer, Cham, 2018, pp. 289--333.

\bibitem{Oliner1959}
A.~A. Oliner and A.~Hessel, ``{Guided Waves on Sinusoidally-Modulated Reactance
  Surfaces},'' \emph{IRE Trans. Antennas Propag.}, vol.~7, no.~5, pp. 201--208,
  1959.

\bibitem{Martin-Moreno2001}
L.~Mart{\'{i}}n-Moreno, F.~J. Garc{\'{i}}a-Vidal, H.~J. Lezec, K.~M. Pellerin,
  T.~Thio, J.~B. Pendry, and T.~W. Ebbesen, ``{Theory of extraordinary optical
  transmission through subwavelength hole arrays},'' \emph{Phys. Rev. Lett.},
  vol.~86, no.~6, pp. 1114--1117, Feb. 2001.

\bibitem{Miyamuru04}
F.~Miyamaru and M.~Hangyo, ``{Finite size effect of transmission property for
  metal hole arrays in subterahertz region},'' \emph{Appl. Phys. Lett.},
  vol.~84, no.~15, pp. 2742--2744, Apr. 2004.

\bibitem{Cao2004}
H.~Cao and A.~Nahata, ``{Resonantly enhanced transmission of terahertz
  radiation through a periodic array of subwavelength apertures},'' \emph{Opt.
  Express}, vol.~12, no.~6, p. 1004, 2004.

\bibitem{Beruete05}
M.~Beruete, M.~Sorolla, I.~Campillo, J.~S. Dolado, L.~Mart{\'{i}}n-Moreno,
  J.~Bravo-Abad, and F.~J. Garc{\'{i}}a-Vidal, ``{Enhanced millimeter wave
  transmission through quasioptical subwavelength perforated plates},''
  \emph{IEEE Trans. Antennas Propag.}, vol.~53, no.~6, pp. 1897--1903, Jun.
  2005.

\bibitem{Medina2009}
F.~Medina, J.~A. Ruiz-Cruz, F.~Mesa, J.~M. Rebollar, J.~R. Montejo-Garai, and
  R.~Marqu{\'{e}}s, ``{Experimental verification of extraordinary transmission
  without surface plasmons},'' \emph{Appl. Phys. Lett.}, vol.~95, no.~7, p.
  071102, 2009.

\bibitem{Kuznetsov2009a}
S.~A. Kuznetsov, M.~Navarro-C{\'{i}}a, V.~V. Kubarev, A.~V. Gelfand,
  M.~Beruete, I.~Campillo, and M.~Sorolla, ``{Regular and anomalous
  extraordinary optical transmission at the THz-gap},'' \emph{Opt. Express},
  vol.~17, no.~14, p. 11730, Jul. 2009.

\bibitem{Beruete2011b}
M.~Beruete, M.~Navarro-C{\'{i}}a, S.~A. Kuznetsov, and M.~Sorolla, ``{Circuit
  approach to the minimal configuration of terahertz anomalous extraordinary
  transmission},'' \emph{Appl. Phys. Lett.}, vol.~98, no.~1, Jan. 2011.

\bibitem{Pendry2004}
J.~B. Pendry, L.~Mart{\'{i}}n-Moreno, and F.~J. Garcia-Vidal, ``{Mimicking
  surface plasmons with structured surfaces.}'' \emph{Science}, vol. 305, no.
  5685, pp. 847--8, 2004.

\bibitem{Hibbins2004}
A.~P. Hibbins, J.~R. Sambles, C.~R. Lawrence, and J.~R. Brown, ``{Squeezing
  millimeter waves into microns},'' \emph{Phys. Rev. Lett.}, vol.~92, no.~14,
  pp. 143\,901--143\,904, Apr. 2004.

\bibitem{Lomakin2006}
V.~Lomakin and E.~Michielssen, ``{Transmission of transient plane waves through
  perfect electrically conducting plates perforated by periodic arrays of
  subwavelength holes},'' \emph{IEEE Trans. Antennas Propag.}, vol.~54, no.~3,
  pp. 970--984, 2006.

\bibitem{Cutler}
C.~Cutler, ``{Genesis of the corrugated electromagnetic surface},'' \emph{Proc.
  IEEE Antennas Propag. Soc. Int. Symp. URSI Natl. Radio Sci. Meet.}, vol.~3,
  pp. 1456--1459, 1994.

\bibitem{Walter1970}
C.~H. Walter, \emph{{Traveling wave antennas}}.\hskip 1em plus 0.5em minus
  0.4em\relax New York: Dover Publications, 1970.

\bibitem{Medina2008}
F.~Medina, F.~Mesa, and R.~Marqu{\'{e}}s, ``{Extraordinary transmission through
  arrays of electrically small holes from a circuit theory perspective},''
  \emph{IEEE Trans. Microw. Theory Tech.}, vol.~56, no.~12, pp. 3108--3120,
  Dec. 2008.

\bibitem{Beruete2011a}
M.~Beruete, M.~Navarro-C{\'{i}}a, and M.~{Sorolla Ayza}, ``{Understanding
  anomalous extraordinary transmission from equivalent circuit and grounded
  slab concepts},'' \emph{IEEE Trans. Microw. Theory Tech.}, vol.~59, no.~9,
  pp. 2180--2188, Sep. 2011.

\bibitem{Munk2000}
{B. A. Munk}, \emph{{Frequency Selective Surfaces: Theory and Design}}.\hskip
  1em plus 0.5em minus 0.4em\relax John Wiley, 2000.

\bibitem{Mittra88}
R.~Mittra, C.~H. Chan, and T.~Cwik, ``{Techniques for analyzing frequency
  selective surfaces--a review},'' \emph{Proc. IEEE}, vol.~76, no.~12, pp.
  1593--1615, Dec. 1988.

\bibitem{Rodriguez-Ulibarri2017}
P.~Rodriguez-Ulibarri, M.~Navarro-Cia, R.~Rodriguez-Berral, F.~Mesa, F.~Medina,
  and M.~Beruete, ``{Annular apertures in metallic screens as extraordinary
  transmission and frequency selective surface structures},'' \emph{IEEE Trans.
  Microw. Theory Tech.}, vol.~65, no.~12, pp. 4933--4946, Dec. 2017.

\bibitem{Camacho2017e}
M.~Camacho, R.~R. Boix, F.~Medina, A.~P. Hibbins, and J.~{Roy Sambles}, ``{On
  the extraordinary optical transmission in parallel plate waveguides for
  non-TEM modes},'' \emph{Opt. Express}, vol.~25, no.~20, p. 24670, Oct. 2017.

\bibitem{Camacho2019b}
M.~Camacho, R.~R. Boix, S.~A. Kuznetsov, M.~Beruete, and M.~Navarro-Cia,
  ``{Far-Field and Near-Field Physics of Extraordinary THz Transmitting
  Hole-Array Antennas},'' \emph{IEEE Trans. Antennas Propag.}, vol.~67, no.~9,
  pp. 6029--6038, 2019.

\bibitem{Camacho2019}
M.~Camacho, A.~P. Hibbins, F.~Capolino, and M.~Albani, ``{Diffraction by a
  truncated planar array of dipoles: A Wiener–Hopf approach},'' \emph{Wave
  Motion}, vol.~89, pp. 28--42, 2019.

\bibitem{Beruete2007b}
M.~Beruete, M.~Navarro-C{\'{i}}a, M.~Sorolla, and I.~Campillo, ``{Polarized
  left-handed extraordinary optical transmission of subterahertz waves},''
  \emph{Opt. Express}, vol.~15, no.~13, p. 8125, Jun. 2007.

\bibitem{Torres2014}
V.~Torres, N.~S{\'{a}}nchez, D.~Etayo, R.~Ortu{\~{n}}o, M.~Navarro-C{\'{i}}a,
  A.~Mart{\'{i}}nez, and M.~Beruete, ``{Compact dual-band terahertz
  quarter-wave plate metasurface},'' \emph{IEEE Photonics Technol. Lett.},
  vol.~26, no.~16, pp. 1679--1682, 2014.

\bibitem{Camacho2016}
M.~Camacho, R.~R. Boix, and F.~Medina, ``{Computationally efficient analysis of
  extraordinary optical transmission through infinite and truncated
  subwavelength hole arrays},'' \emph{Phys. Rev. E - Stat. Nonlinear, Soft
  Matter Phys.}, vol.~93, no.~6, p. 063312, Jun. 2016.

\bibitem{Harrington93}
R.~F. Harrington, \emph{{Field computation by moment methods}}.\hskip 1em plus
  0.5em minus 0.4em\relax New York, USA: Wiley-IEEE Press, 1993.

\bibitem{Camacho2017b}
M.~Camacho, R.~R. Boix, and F.~Medina, ``{Comparative study between resonant
  transmission and extraordinary transmission in truncated periodic arrays of
  slots},'' in \emph{2017 IEEE MTT-S Int. Conf. Numer. Electromagn.
  Multiphysics Model. Optim. RF, Microwave, Terahertz Appl.}\hskip 1em plus
  0.5em minus 0.4em\relax IEEE, May 2017, pp. 257--259.

\bibitem{Lerer93}
A.~M. Lerer and A.~G. Schuchinsky, ``{Full-wave analysis of three-dimensional
  planar structures},'' \emph{IEEE Trans. Microw. Theory Techn.}, vol.~41,
  no.~11, pp. 2002--2015, 1993.

\bibitem{Camacho2019a}
M.~Camacho, R.~R. Boix, and F.~Medina, ``{NUFFT for the Efficient Spectral
  Domain MoM Analysis of a Wide Variety of Multilayered Periodic Structures},''
  \emph{IEEE Trans. Antennas Propag.}, vol.~67, no.~10, pp. 6551--6563, Jun.
  2019.

\bibitem{Camacho2017a}
M.~Camacho, R.~R. Boix, F.~Medina, A.~P. Hibbins, and J.~R. Sambles,
  ``{Theoretical and experimental exploration of finite sample size effects on
  the propagation of surface waves supported by slot arrays},'' \emph{Phys.
  Rev. B}, vol.~95, no.~24, p. 245425, Jun. 2017.

\bibitem{Yakovlev1997}
A.~B. Yakovlev and G.~W. Hanson, ``{On the nature of critical points in leakage
  regimes of a conductor-backed coplanar strip line},'' \emph{IEEE Trans.
  Microw. Theory Tech.}, vol.~45, no.~1, pp. 87--94, 1997.

\bibitem{Camacho2019c}
M.~Camacho, R.~R. Boix, F.~Medina, A.~P. Hibbins, and J.~R. Sambles,
  ``{Extraordinary Transmission and Radiation from Finite by Infinite Arrays of
  Slots},'' \emph{IEEE Trans. Antennas Propag.}, vol.~68, no.~1, pp. 581--586,
  2019.

\bibitem{Freer2019}
S.~Freer, M.~Camacho, S.~A. Kuznetsov, R.~R. Boix, M.~Beruete, and
  M.~Navarro-C{\'{i}}a, ``{Revealing the Underlying Mechanisms Behind TE
  Extraordinary THz Transmission},'' \emph{Photonics Res.}, vol.~8, no.~4, pp.
  430--439, 2020.

\bibitem{Navarro-Cia2018a}
M.~Navarro-C{\'{i}}a, V.~Pacheco-Pe{\~{n}}a, S.~A. Kuznetsov, and M.~Beruete,
  ``{Extraordinary THz Transmission with a Small Beam Spot: The Leaky Wave
  Mechanism},'' \emph{Adv. Opt. Mater.}, vol.~6, no.~8, p. 1701312, feb 2018.

\bibitem{Worldwide2010}
Aerotech, \emph{{Aerotech PRO165LM Series Stage User's Manual P/N: EDS142
  (Revision 1.06.00)}}, 2010.

\bibitem{Penchev2017}
P.~Penchev, S.~Dimov, D.~Bhaduri, S.~L. Soo, and B.~Crickboom, ``{Generic
  software tool for counteracting the dynamics effects of optical beam delivery
  systems},'' \emph{Proc. Inst. Mech. Eng. Part B J. Eng. Manuf.}, vol. 231,
  no.~1, pp. 48--64, Jan. 2017.

\end{thebibliography}
\bibliographystyle{IEEEtran}

 \end{document}